\documentclass[twocolumn]{aastex62}

\usepackage{graphicx}
\usepackage{amsmath}	
\usepackage{amssymb}
\usepackage{booktabs}
\usepackage{enumerate}
\usepackage{dirtytalk}
\usepackage{float}
\usepackage{enumitem}
\usepackage{booktabs}
\usepackage{lipsum} 
\usepackage{float}
\usepackage{xcolor}
\usepackage{multirow}
\usepackage{makecell}
\usepackage[english]{babel}
\shorttitle{Interior and Atmosphere of K2-18b}
\shortauthors{Madhusudhan et al.}

\begin{document}

\title{The interior and atmosphere of the habitable-zone exoplanet K2-18b}

\author{Nikku Madhusudhan}
\email{E-mail: nmadhu@ast.cam.ac.uk}
\affil{Institute of Astronomy, University of Cambridge, Madingley Road, Cambridge CB3 0HA, UK}

\author{Matthew C. Nixon}
\affil{Institute of Astronomy, University of Cambridge, Madingley Road, Cambridge CB3 0HA, UK}

\author{Luis Welbanks}
\affil{Institute of Astronomy, University of Cambridge, Madingley Road, Cambridge CB3 0HA, UK}

\author{Anjali A. A. Piette}
\affil{Institute of Astronomy, University of Cambridge, Madingley Road, Cambridge CB3 0HA, UK}

\author{Richard A. Booth}
\affil{Institute of Astronomy, University of Cambridge, Madingley Road, Cambridge CB3 0HA, UK}

\begin{abstract}
Exoplanets orbiting M dwarfs present a valuable opportunity for their detection and atmospheric characterisation. This is evident from recent inferences of H$_2$O in such atmospheres, including that of the habitable-zone exoplanet K2-18b. With a bulk density between Earth and Neptune, K2-18b may be expected to possess a H/He envelope. However, the extent of such an envelope and the thermodynamic conditions of the interior remain unexplored. In the present work, we investigate the atmospheric and interior properties of K2-18b based on its bulk properties and its atmospheric transmission spectrum. We constrain the atmosphere to be H$_2$-rich with a H$_2$O volume mixing ratio of $0.02-14.8$\%, consistent with previous studies, and find a depletion of CH$_4$ and NH$_3$, indicating chemical disequilibrium. We do not conclusively detect clouds/hazes in the observable atmosphere. We use the bulk parameters and retrieved atmospheric properties to constrain the internal structure and thermodynamic conditions in the planet. The constraints on the interior allow multiple scenarios between rocky worlds with massive H/He envelopes and water worlds with thin envelopes. We constrain the mass fraction of the H/He envelope to be $\lesssim 6$\%; spanning $\lesssim 10^{-5}$ for a predominantly water world to $\sim6$\% for a pure iron interior. The thermodynamic conditions at the surface of the H$_2$O layer range from the super-critical to liquid phases, with a range of solutions allowing for habitable conditions on K2-18b. Our results demonstrate that the potential for habitable conditions is not necessarily restricted to Earth-like rocky exoplanets.
\end{abstract}


\section{Introduction}
\label{sec:intro}

\begin{figure*}
\centering
\includegraphics[width=1.0\textwidth]{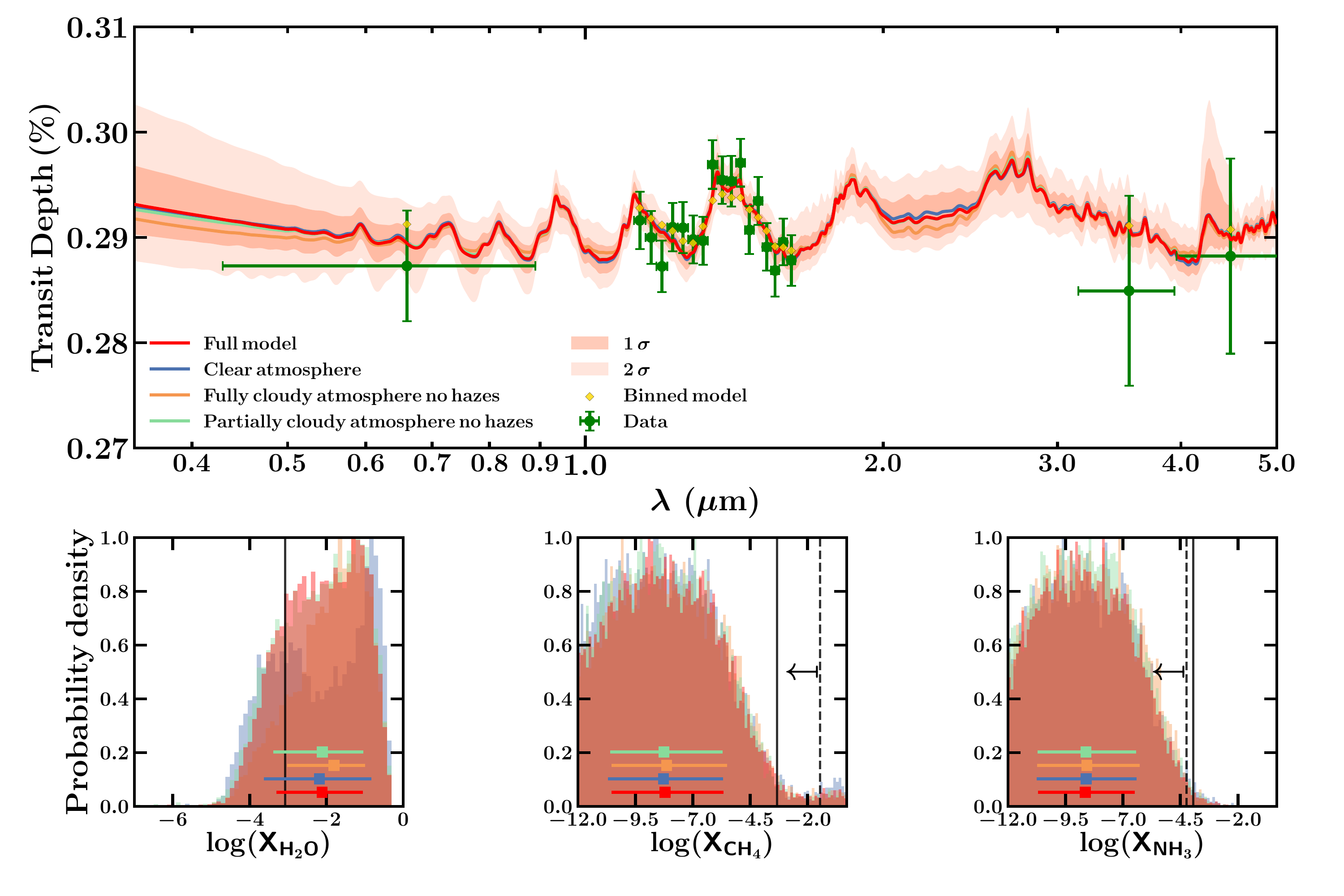}
    \caption{Atmospheric retrieval from the transmission spectrum of K2-18b. 
    \textit{Top}: Observations (green) and retrieved model spectra for the four different model considerations in Table ~\ref{table:atmos}. Shaded regions represent 1$\sigma$ and 2$\sigma$ confidence intervals for the full model, with yellow points showing the model binned to the data resolution. The observations were adopted from \citet{benneke2019}. 
    \textit{Bottom}: Posterior distributions for the retrieved volume mixing ratios of H$_2$O, CH$_4$, and NH$_3$. The 99\% upper limits for the full model on CH$_4$ and NH$_3$ are shown by the arrows and dashed lines. Equilibrium solar values are shown by solid black lines.}
    \label{fig:atmos}
\end{figure*}

Recent exoplanet detection surveys have revealed high occurrence rates of low-mass planets orbiting M Dwarfs \citep{dressing2015,mulders2015}. The low masses, sizes and temperatures of M Dwarfs also mean that the planet-star contrast is favourable for planetary detection and characterisation. This `small-star opportunity' has led to several detections of low-mass planets ($<10M_{\oplus}$) in the habitable-zones of M Dwarf hosts such as Trappist-1 \citep{gillon2017}, Proxima Centauri \citep{anglada-escude2016}, K2-18 \citep{montet2015,foreman-mackey2015}, and LHS~1140 \citep{dittmann2017}. 

The habitable-zone transiting exoplanet K2-18b is a particularly good example  \citep{foreman-mackey2015,montet2015}. The close proximity and small size of its host star make precise measurements of the planetary mass, radius, and atmospheric spectra viable \citep{benneke2017,cloutier2019}, as exemplified by the recent detection of H$_2$O in its atmosphere \citep{tsiaras2019,benneke2019}. The habitable-zone temperature of K2-18b provides further impetus for detailed characterisation of its interior and atmosphere.

Given its mass ($M_p$ = 8.63~$\pm$~1.35 $M_{\oplus}$, \citealt{cloutier2019}) and radius ($R_p$ = 2.610~$\pm$~0.087 $R_{\oplus}$, \citealt{benneke2019}), K2-18b has a bulk density (2.67$^{+0.52}_{-0.47}$~g/cm$^3$, \citealt{benneke2019}). This density, between that of Earth and Neptune, may be thought to preclude a purely rocky or icy interior and require a hydrogen-rich outer envelope. However, the  extent of such an envelope and the conditions at the interface between the envelope and the underlying interior have not been explored. We note that the mass and radius of the planet have recently been revised \citep{benneke2019}, which may have impacted inferences made using previous values \citep{cloutier2017,tsiaras2019}. 

 Previous studies of planets with similar masses and radii, such as GJ~1214b, suggested envelope mass fractions $\lesssim$7\% \citep{rogers2010,nettelmann2011,valencia2013}. GJ~1214b is expected to host super-critical H$_2$O below the envelope at pressures and temperatures too high to be conducive for life \citep{rogers2010}. However, while GJ~1214b has an equilibrium temperature ($T_{eq}$) of $\sim500$~K, K2-18b may be more favourable given its lower $T_{eq}\sim250-300$~K. 

In the present work, we conduct a systematic study to constrain both the  atmospheric and interior composition of K2-18b based on extant data along with detailed atmospheric retrievals and internal structure models.

\begin{deluxetable*}{c|ccccc}
\tablecaption{Retrieved atmospheric properties from the transmission spectrum of K2-18b.  \label{table:atmos}}
\tablecolumns{6}

\tablewidth{0pt}
\tablehead{
\colhead{Model} & \colhead{$\log$(X$_{\text{H}_2\text{O}}$)} & \colhead{$\log$(X$_{\text{CH}_4}$)} & \colhead{$\log$(X$_{\text{NH}_3}$)}  & \colhead{$\ln$($\mathcal{Z}$)} & \colhead{ Detection Significance (DS)} 
}
\startdata
Case 1: Full model,  inhomogenous clouds and hazes & $-2.11 ^{+ 1.06 }_{- 1.19 }$ & $-8.20 ^{+ 2.53 }_{- 2.34 }$ & $-8.64 ^{+ 2.15 }_{- 2.06 }$ & $179.15$	& 	Reference \\
No H$_2$O & N/A & $-1.11 ^{+ 0.53 }_{- 1.22 }$ & $-7.27 ^{+ 2.91 }_{- 2.92 }$ &  $175.30$& 	$3.25$ \\
Case 2: Clear atmosphere & $-2.18 ^{+ 1.35 }_{- 1.44 }$	& 	$-8.27 ^{+ 2.59 }_{- 2.42 }$ & $-8.60 ^{+ 2.19 }_{- 2.16 }$ 	& $179.05$	& 	$1.20$\\
Case 3: Opaque cloud deck 	& 	$-1.80 ^{+ 0.81 }_{- 1.22 }$	& 	$-8.13 ^{+ 2.64 }_{- 2.41 }$ & 	$-8.57 ^{+ 2.30 }_{- 2.17 }$   & $179.09$	&  $1.06$ \\
Case 4: Inhomogenous clouds &	$-2.10 ^{+ 1.07 }_{- 1.28 }$ &	$-8.26 ^{+ 2.56 }_{- 2.34 }$ &	$-8.61 ^{+ 2.18 }_{- 2.10 }$ 	& 	$179.41$	& 	N/A 
\enddata
\tablecomments{Four models are considered with different treatments of clouds and hazes. For each model, the volume mixing ratios ($\log$(X$_{\text{H}_2\text{O}}$), $\log$(X$_{\text{CH}_4}$), and $\log$(X$_{\text{NH}_3}$)) are shown along with the Bayesian  evidence ($\ln$($\mathcal{Z}$)) and detection significance (DS). The DS is derived from the Bayesian evidence and a value below 2.0$\sigma$ is considered weak \citep{trotta2008}. The preference of the reference model (case 1) over other models is quantified by the DS. For example, the DS for case 2 implies that case 1 is preferred over case 2 at 1.2$\sigma$. H$_2$O is detected at 3.25$\sigma$ and clouds/hazes at only $\sim$1$\sigma$.}
\end{deluxetable*}

\section{Atmospheric Properties}
\label{sec:atmosphere}

We retrieve the atmospheric properties of K2-18b using its broadband transmission spectrum reported by \citet{benneke2019}. The data include observations from the HST WFC3~G141 grism (1.1-1.7 $\mu$m), photometry in the Spitzer IRAC 3.6 $\mu$m and 4.5 $\mu$m bands, and optical photometry in the K2 band (0.4-1.0 $\mu$m). We perform the atmospheric retrieval using an adaptation of the AURA retrieval code \citep{pinhas2019, welbanks2019}. Our model solves line-by-line radiative transfer in a plane-parallel atmosphere in transmission geometry. The model  assumes hydrostatic equilibrium and considers prominent opacity sources in the observed spectral bands as well as homogeneous/inhomogeneous cloud/haze coverage. Clouds are included through a gray cloud deck with cloud-top pressure ($\text{P}_\text{c}$) as a free parameter. Hazes are included as a modification to Rayleigh-scattering through parameters for the scattering slope ($\gamma$) and a Rayleigh-enhancement factor ($a$). The opacity sources include H$_2$O \citep{rothman2010}, CH$_4$ \citep{yurchenko2014}, NH$_3$ \citep{yurchenko2011}, CO$_2$ \citep{rothman2010}, HCN \citep{barber2014}, and collision-induced absorption due to H$_2$-H$_2$ and H$_2$-He \citep{richard2012}. 

The model comprises 16 free parameters: abundances of 5 molecules, 6 parameters for the pressure-temperature ($P$-$T$) profile, 4 cloud/haze parameters, and 1 parameter for the reference pressure $P_{\text{ref}}$ at $R_p$ \citep[e.g.,][]{welbanks2019b}. The Bayesian parameter estimation is conducted using the Nested Sampling algorithm MultiNest \citep{feroz2009} through PyMultiNest \citep{Buchner2014}. We conduct retrievals for four model configurations: (1) a full model including inhomogeneous clouds and hazes, (2) a clear atmosphere, (3) an atmosphere with an opaque cloud deck but no hazes, and (4) an atmosphere with inhomogeneous clouds but no hazes. The atmospheric constraints are shown in Figure~\ref{fig:atmos} and Table ~\ref{table:atmos}. 

We confirm the high-confidence detection of H$_2$O in a H$_2$-rich atmosphere as reported by \citet{benneke2019} and \citet{tsiaras2019}. Our abundance estimates are consistent to within 1$\sigma$ between all four model configurations and with \citet{benneke2019}. The derived H$_2$O volume mixing ratio ranges between  0.02-14.80\%, with median values of 0.7-1.6\% between the 4 model cases, as shown in Table~\ref{table:atmos}. The case with an opaque cloud deck (a clear atmosphere) retrieves slightly higher (lower) H$_2$O abundances as expected \citep{welbanks2019}. Our derived H$_2$O abundance range corresponds to an O/H ratio of 0.2-176.8$\times$solar, assuming all the oxygen is in H$_2$O as expected in H$_2$-rich atmospheres at such low temperatures \citep{burrows1999}. The median H$_2$O abundance is 9.3$\times$solar for the full model, case 1. We cannot compare our results with \citet{tsiaras2019} as their retrievals were based on only the HST WFC3 data and used older measurements of the planetary mass and radius which could have biased their inferences.

We find a depletion of CH$_4$ and NH$_3$ in the atmosphere. For a H$_2$-rich atmosphere at $\sim$300 K, CH$_4$ and NH$_3$ are expected to be dominant carriers of carbon and nitrogen, respectively, in chemical equilibrium \citep{burrows1999}, as also seen for the gas and ice giants in the solar system \citep{atreya2016}. Assuming solar elemental ratios (i.e., C/O = 0.55, N/O = 0.14), the CH$_4$/H$_2$O (NH$_3$/H$_2$O) ratio is expected to be $\sim$0.5 ($\sim$0.1). However, we do not detect CH$_4$ or NH$_3$ despite their strong absorption in the HST WFC3 and/or Spitzer 3.6 $\mu$m bands. As shown in Figure~\ref{fig:atmos}, the retrieved posteriors of the CH$_4$ and NH$_3$ abundances are largely sub-solar, with 99\% upper limits of 3.47$\times 10^{-2}$ and 5.75$\times 10^{-5}$, respectively. These sub-solar values are in contrast to the largely super-solar H$_2$O, arguing against chemical equilibrium at solar elemental ratios.

We do not find strong evidence for clouds/hazes in the atmosphere. Our model preference for clouds/hazes, relative to the cloud-free case, is marginal (1.2$\sigma$) compared to \citet{benneke2019} (2.6$\sigma$). Our retrieved cloud-top pressure ($P_c$) for the full case is weakly constrained to 0.1 mbar to 2 bar, close to the observable photosphere. Finally, we retrieve  $P_{\text{ref}}$ for the full case to be $12 - 174$ mbar corresponding to $R_p$. The median value of 0.05 bar is used as the surface boundary condition, pressure $P_0$, for the internal structure models in section~\ref{sec:interior_model}. 

\begin{figure}
\includegraphics[width=\linewidth]{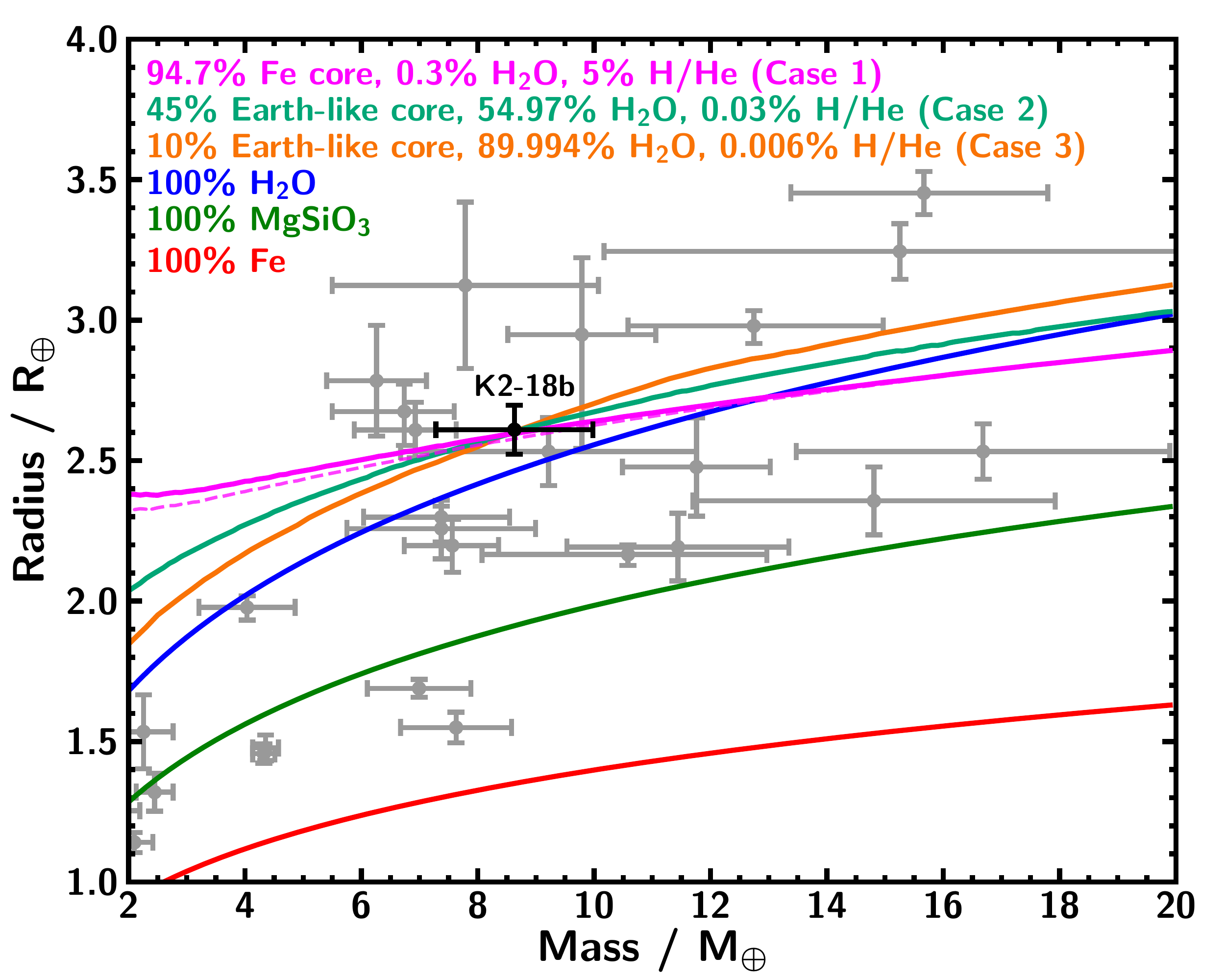}
    \caption{Model mass-radius (M-R) relations for planets with different compositions. The mass fractions are shown in the legend. The solid magenta, teal, and orange curves show cases with three representative compositions, discussed in section~\ref{sec:scenarios}, that all fit the mass and radius of K2-18b equally well. The dashed magenta line represents the same composition as the solid magenta line, but with a mixed H$_2$O-H/He envelope. Also shown are exoplanets whose masses and radii are known to $\geq 3\sigma$ with $T_{eq} <$~1000~K, from TEPCat \citep{southworth2011}.}
    \label{fig:mr_diagram}
\end{figure}

\begin{figure*}
\centering
\begin{center}$
\begin{array}{cc}
\includegraphics[width=0.5\textwidth]{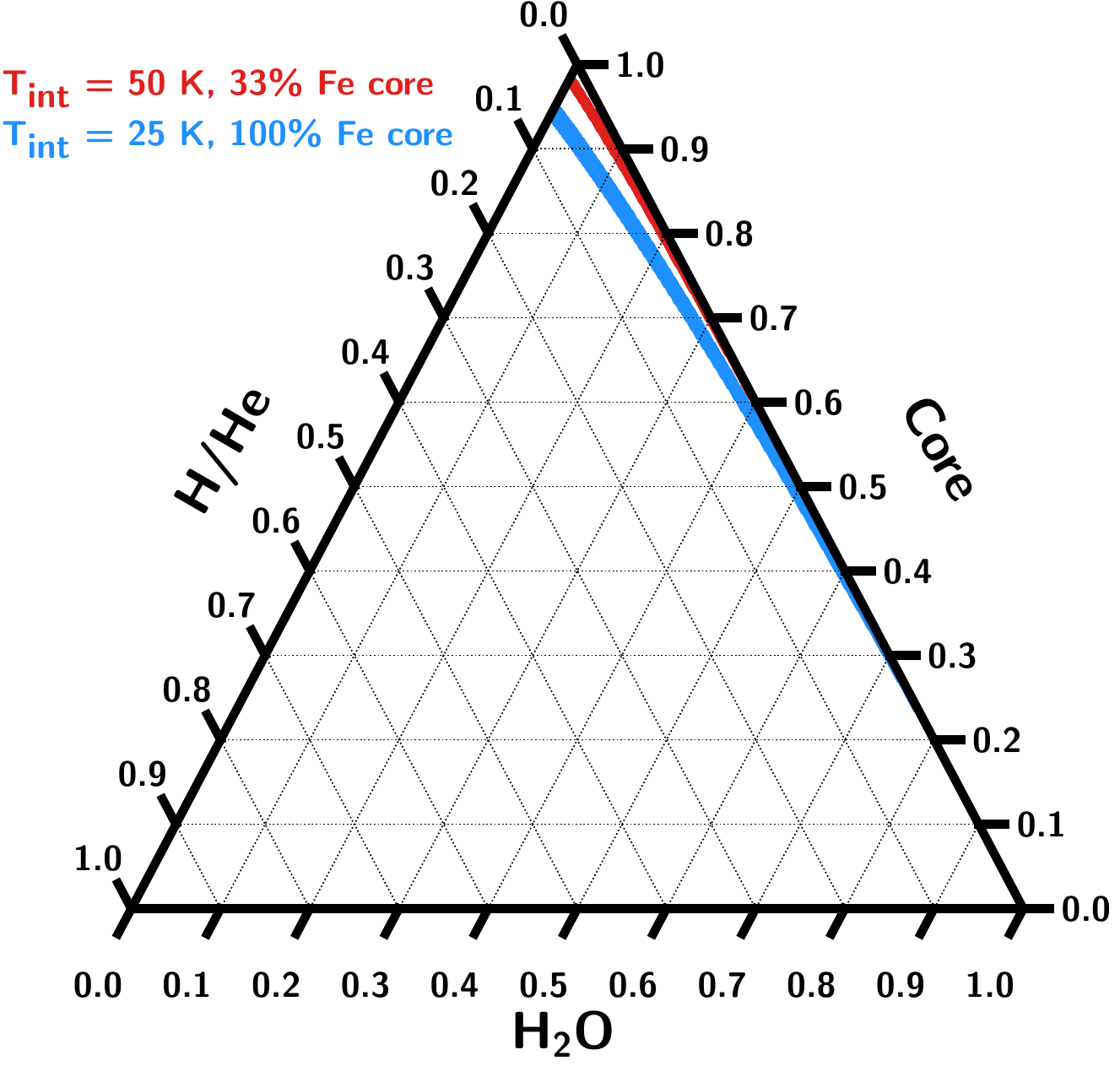}
\includegraphics[width=0.5\textwidth]{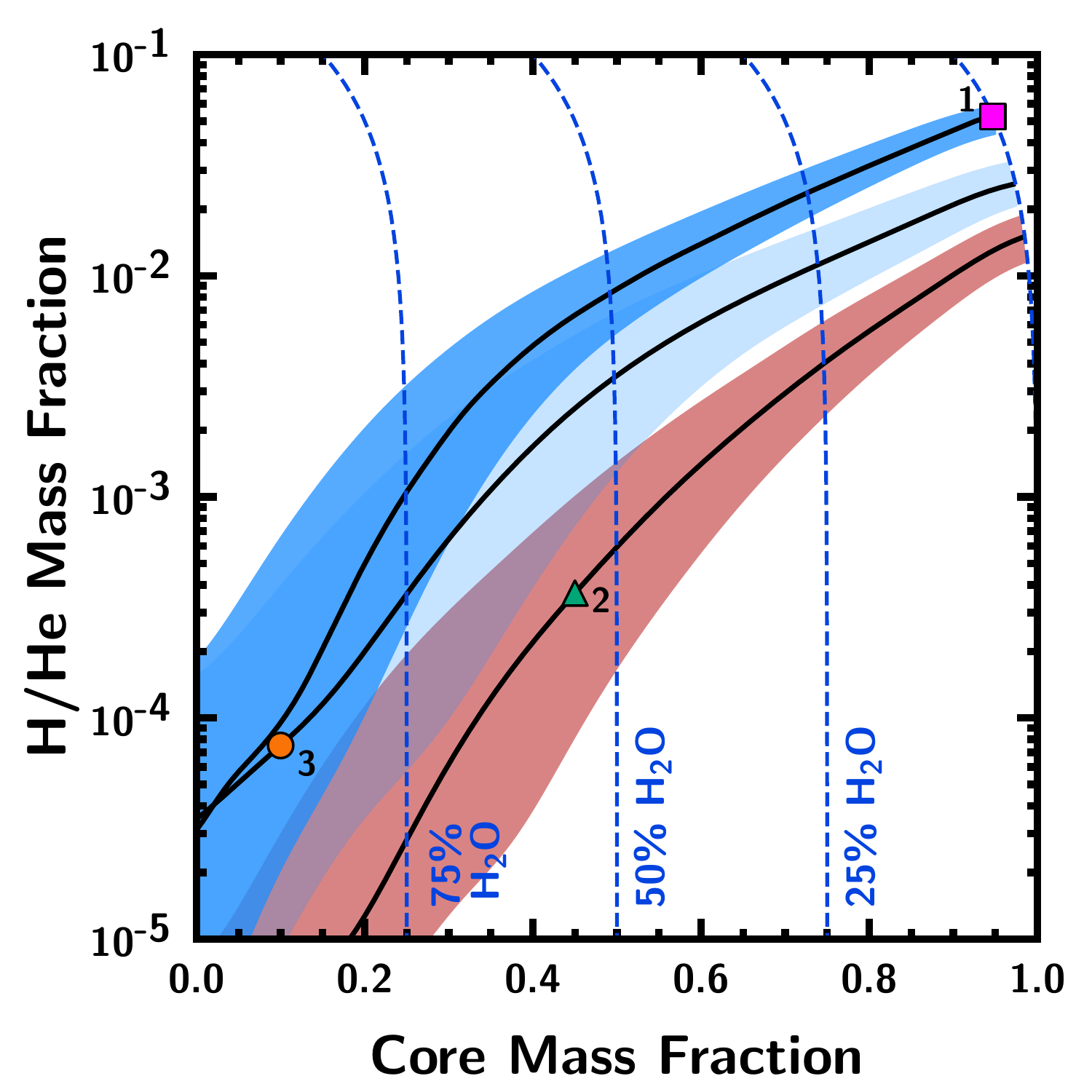}
\end{array}$
\end{center}
    \caption{\textit{Left}: Ternary diagram showing best-fitting ($\leq1\sigma$) interior compositions allowed by the mass and radius of K2-18b for two end-member core compositions and interior temperatures.   \textit{Right}: Envelope vs core mass fraction for model solutions. The dark red and blue shaded regions show the same cases as in the ternary diagram. The pale blue region shows an additional case with $T_{\rm int} = 25K$ and an Earth-like core for comparison. The black lines in each case show the loci of the best-fit solutions. The magenta square, teal triangle and orange circle represent the rocky world, intermediate and water world scenarios discussed in section \ref{sec:discussion}. The H/He mass fraction ($x_{\rm env}$) is constrained to be $<$3.3\% ($<$6.2\%) for models with an Earth-like (33\% Fe) core and a pure (100\%) Fe core, respectively.}
    \label{fig:ternary}
\end{figure*}

\section{Internal Structure and Composition}
\label{sec:interior}

In this section we use the observed bulk properties of K2-18b, namely the planetary mass ($M_p$), radius ($R_p$), and its atmospheric properties, to constrain its internal structure and thermodynamic conditions. 

\subsection{Internal structure model}
\label{sec:interior_model}

We model the interior of the planet with a canonical four-layer structure. The model comprises a two-component Fe+rock core consisting of an inner Fe layer and an outer silicate layer, a layer of H$_2$O, and an outer H/He envelope. Such a model spans the possible internal structures and compositions of super-Earths and mini-Neptunes \citep[e.g.][]{valencia2010,valencia2013, rogers2011,lopez2014}, as well as terrestrial planets and ice giants in the solar system \citep{guillot2014}. The mass fractions of the different components ($x_{\rm Fe}$, $x_{\rm rock}$, $x_{\rm H_2O}$,  $x_{\rm env}$) are free parameters in the model and sum to unity. Our present model is adapted from a three-layer model for super-Earths from \citet{madhu2012} comprising of Fe, rock, and H$_2$O, with the H/He envelope added in the present work.

The model solves the standard internal structure equations of hydrostatic equilibrium and mass continuity assuming spherical symmetry. The equation of state (EOS) for each of the two inner layers is adopted from \citet{seager2007} who use the Birch-Murnaghan EOS \citep{birch1952} for Fe \citep{ahrens2000} and MgSiO$_3$ perovskite \citep{karki2000}. For the H$_2$O layer we use the temperature-dependent H$_2$O EOS compiled by \citet{thomas2016} from \citet{french2009,sugimura2010,fei1993,seager2007} and \citet{wagner2002}. For the gaseous envelope we use the latest H/He EOS from \citet{chabrier2019} for a solar helium mass fraction ($Y = 0.275$).

The EOS in the H/He and H$_2$O layers can have a significant temperature dependence which we consider in our model. Past studies \citep{rogers2011,valencia2013} considered analytic $P$-$T$ profiles for irradiated atmospheres derived using double gray approximations \citep{hansen2008,guillot2010} with the internal and external fluxes and opacities as free parameters. We calculate self-consistent dayside $P$-$T$ profiles for K2-18b in the H/He envelope using the GENESIS code \citep{gandhi2017}. GENESIS solves line-by-line radiative transfer under assumptions of hydrostatic, radiative-convective and thermochemical equilibrium. We include opacity due to H$_2$O \citep{rothman2010}, as detected in the transmission spectrum (section \ref{sec:atmosphere}), H$_2$ Rayleigh scattering, clouds and H$_2$-H$_2$ and H$_2$-He collision-induced absorption. We use a H$_2$O abundance of 10$\times$solar (see section \ref{sec:atmosphere}) and also use 10$\times$solar abundances for the cloud species. We include KCl, ZnS and Na$_2$S clouds \citep{morley2013}, for which we obtain opacities from \citet{pinhas2017}. We further include water ice clouds using opacities from \citet{Budaj2015}. 
 
The $P$-$T$ profile also depends on the planetary internal flux, which is characterised by the internal temperature $T_\mathrm{int}$. We consider values of $T_\mathrm{int}$ which span the range expected for a planet with the mass and radius of K2-18b and an age of $1-10$ Gyr, with envelope compositions from solar to water-rich. We choose end-member cases of $T_\mathrm{int}=25\text{K}$ and $50\text{K}$, consistent with previous studies on planets of similar mass and radius, e.g., GJ~1214b \citep[e.g.,][]{valencia2013}. The GENESIS models are calculated between pressures of $10^{-5}-10^3$ bar, and assume full redistribution of the incident stellar irradiation. We explore a range of $P$-$T$ profiles and choose two representative cases, with different $T_\mathrm{int}$, discussed further in sections~\ref{sec:interior_constraints} and \ref{sec:hhb}. Where required by the internal structure model, the bottom of the $P$-$T$ profile of the H/He envelope is continued to deeper pressures using the adiabatic gradient from \citet{chabrier2019}. We also employ an adiabatic temperature profile in the H$_2$O layer, following \citet{thomas2016}.

\subsection{Constraints on interior composition}
\label{sec:interior_constraints}

Figure~\ref{fig:mr_diagram} shows mass-radius relations for models with different interior compositions. We explore the full range of plausible interior compositions in three components: $x_{\rm core} = x_{\rm Fe} + x_{\rm rock}$, $x_{\rm H_2O}$, and $x_{\rm env}$, where $x_i$ = $M_i/M_p$ is the mass fraction of each component $i$. For each atmospheric $P$-$T$ profile considered, we explore two different core compositions: (1) an Earth-like core made of 33\% Fe, 67\% rock by mass, and (2) a pure Fe core, the densest possible composition. Here, we discuss results from two end-member cases: (1) a pure Fe core with $T_{\rm int} = 25$K, and (2) an Earth-like (33\% Fe) core with $T_{\rm int} = 50$K. Solutions for all other cases lie between these two cases. 

As shown in Figure~\ref{fig:ternary}, while a wide range of core and H$_2$O mass fractions are permitted, we place a stringent upper limit on the mass fraction of the H/He envelope: $x_{\rm env} = 6.2$\%. This maximal $x_{\rm env}$ corresponds to the case of a pure Fe core, with $x_{\rm core}\sim94$\%, underlying the H/He envelope with no $x_{\rm H_2O}$; here it is assumed that the atmospheric H$_2$O is not mixed in the envelope. However, if the retrieved atmospheric H$_2$O abundance is assumed to be well mixed in the envelope then the maximal $x_{\rm env} = 6$\% with $x_{\rm H_2O}=0.4$\% by mass; low, but still significantly higher than that of the Earth's oceans ($\sim$0.02\%). %

We find that a substantial gaseous H/He envelope is not necessary to explain the density of K2-18b. Figure~\ref{fig:ternary} shows the $x_{\rm env}$ required for different $x_{\rm core}$. At one extreme, a $\sim$100\% H$_2$O interior with no rocky core can explain the data with an $x_{\rm env}$ of just $\sim$10$^{-6}$, comparable to the mass fraction of the Earth's atmosphere. The presence of a rocky core would necessitate at least a thin H/He envelope. However, even considering a reasonable $x_{\rm core} = 10-50$\% still requires $x_{\rm env}$ of only $\sim 10^{-5}-10^{-2}$, as shown in Figure~\ref{fig:ternary}. Model solutions with the hotter $P$-$T$ profile and/or lower Fe content in the core require smaller $x_{\rm env}$ for a given $x_{\rm core}$.

We have also considered models with miscible H$_2$O and H/He envelopes. We follow the approach of \citet{soubiran2015}, using an additive volume law for mixtures. Assuming that the median H$_2$O mixing ratio in the atmosphere is representative of the mixed (H$_2$O-H/He) envelope, we find that the difference in radius between the mixed and non-mixed models is less than half of the measured uncertainty (see Figure~\ref{fig:mr_diagram}). The constraint on the envelope mass fraction from this mixed case is $x_{\rm env} = 2.5-6.4$\%, consistent with, and a subset of, the constraints discussed above. Note that in this case $x_{\rm env}$ includes both the H/He and H$_2$O mass fraction. 

\begin{figure*}
\centering
\begin{center}$
\begin{array}{cc}
\includegraphics[width=0.5\textwidth]{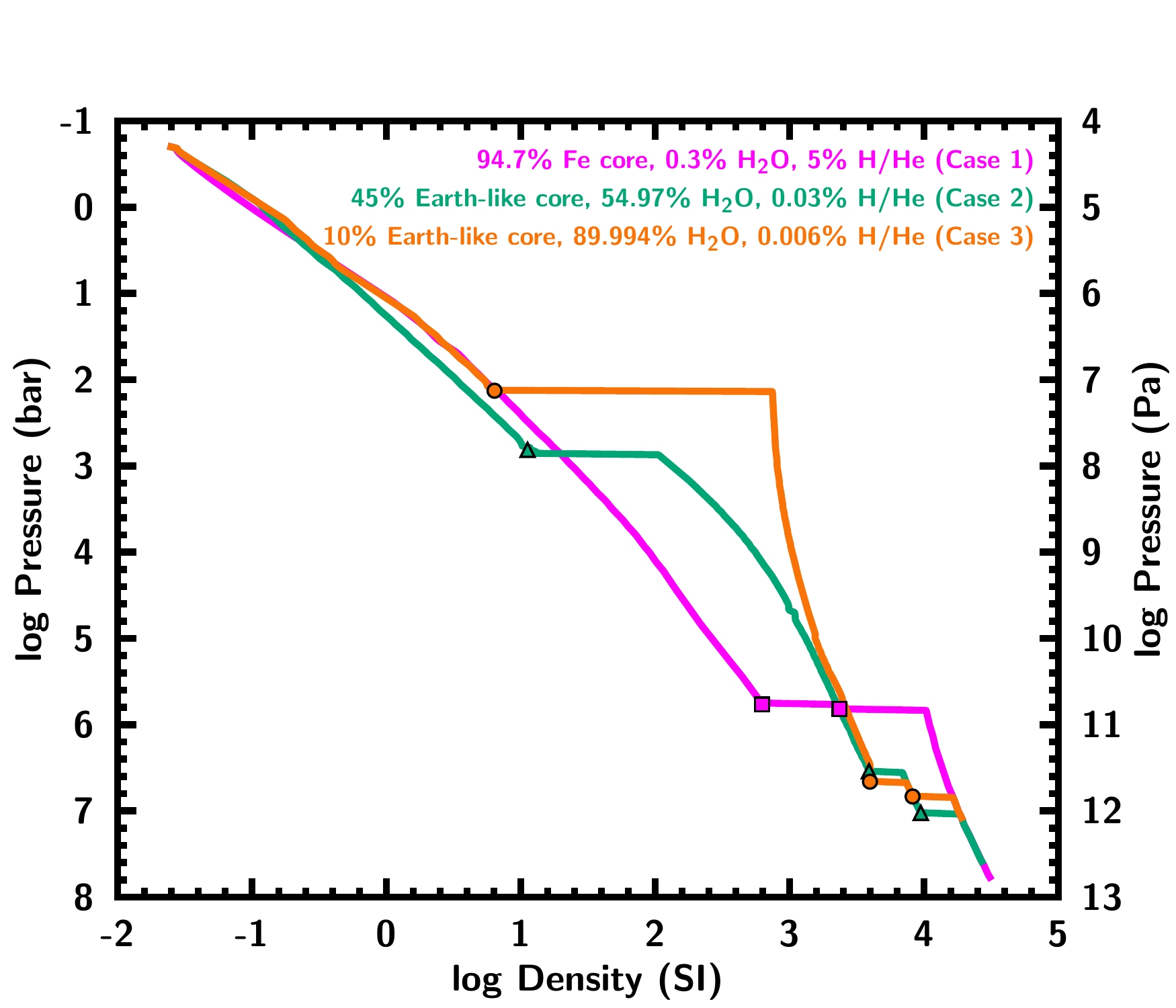}
\includegraphics[width=0.5\textwidth]{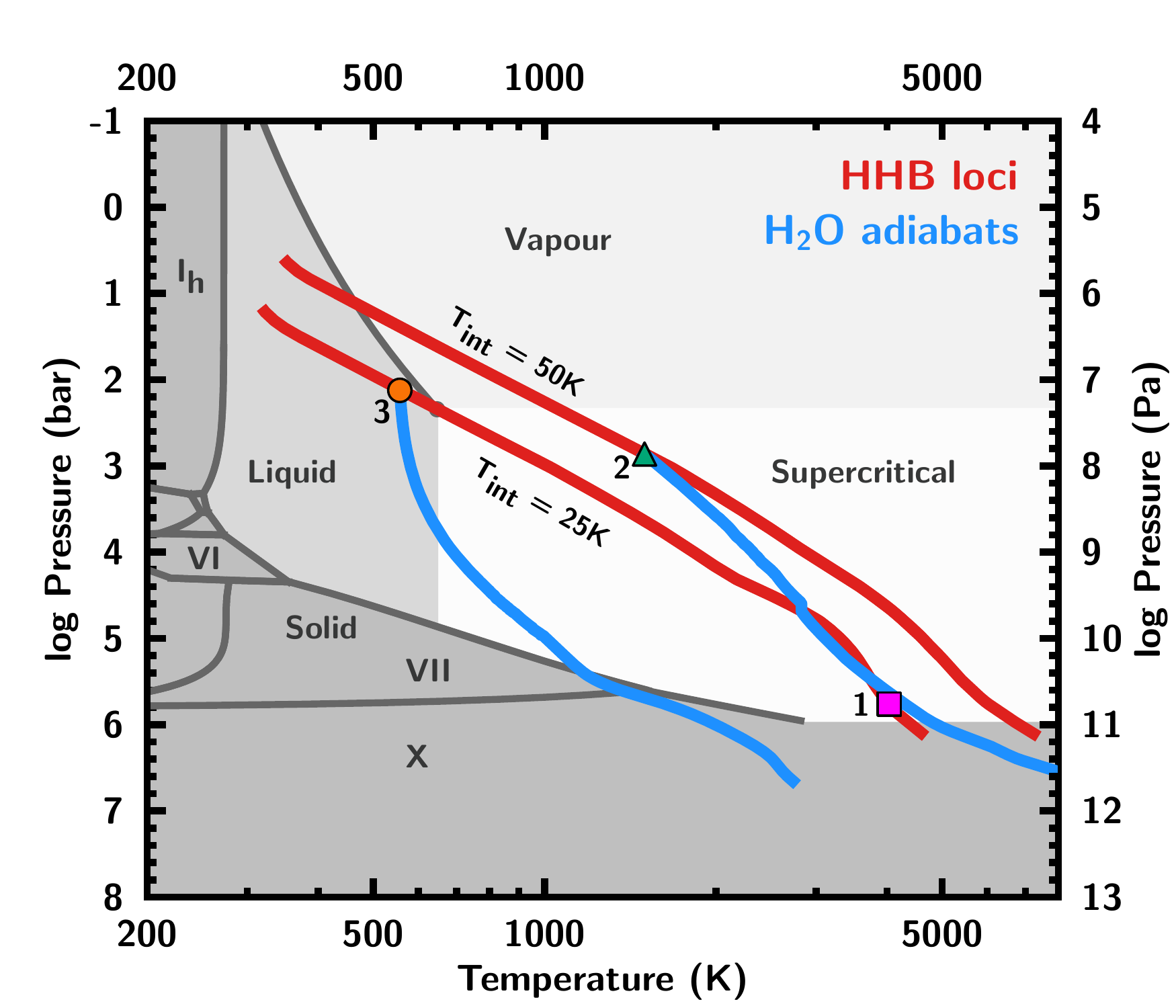}
\end{array}$
\end{center}
    \caption{\textit{Left}: Pressure-density profiles for three possible compositions of K2-18b discussed in Section \ref{sec:discussion}. The transitions between components are marked. \textit{Right}: Thermodynamic conditions at the H$_2$O-H/He boundary (HHB). The red lines indicate the range of possible pressures and temperatures at the HHB for two values of $T_{\rm int}$ considered. They trace the two model $P$-$T$ profiles in the H/He layer. For each case the models span both Earth-like and Fe-only core compositions. The phase diagram of H$_2$O is shown in the background. The square, circle and triangle correspond to the representative cases from the left-hand panel with the same color. We only show solutions with reasonable core mass fractions ($\geq$10\%); less massive cores lead to lower $P$ and $T$ at the HHB. The blue lines show the adiabatic temperature profiles in the H$_2$O layer below the HHB for the three examples.}
    \label{fig:phase_diagram}
\end{figure*}

\subsection{Atmosphere-Ocean Boundary}
\label{sec:hhb}

 Our constraints on the interior compositions of K2-18b result in a wide range of thermodynamic conditions at the H$_2$O-H/He boundary (HHB). The pressure ($P_{\rm HHB}$) and Temperature ($T_{\rm HHB}$) at the HHB for the model solutions are shown in Figure~\ref{fig:phase_diagram}. Each point on the HBB loci denotes the transition from the $P$-$T$ profile in the H/He envelope to the corresponding H$_2$O adiabat. The $P_{\rm HHB}$ and $T_{\rm HHB}$ depend on the H/He envelope mass fraction. For a given $P$-$T$ profile, larger envelopes result in higher $P_{\rm HHB}$ and $T_{\rm HHB}$. For example, solutions with $x_{\rm env} \gtrsim 1$\% lead to $P_{\rm HHB}$ and $T_{\rm HHB}$ corresponding to the super-critical phase of H$_2$O. As shown in Figure~\ref{fig:ternary}, solutions with higher $x_{\rm env}$ correspond to higher $x_{\rm core}$ and lower $x_{\rm H_2O}$. 

Conversely, solutions with lower $x_{\rm core}$, and hence lower $x_{\rm env}$ and higher $x_{\rm H_2O}$, lead to lower $P_{\rm HHB}$ and $T_{\rm HHB}$ with H$_2$O in vapour or liquid phases at the HHB. For example, an $x_{\rm core}\lesssim$~30\% leads to a $P_{\rm HHB}$ and $T_{\rm HHB}$ corresponding to the liquid phase of H$_2$O, for the cooler $P$-$T$ profile (with $T_{\rm int}=25$K). For $x_{\rm core}\sim$10\% or less, the $P_{\rm HHB}$ and $T_{\rm HHB}$ approach STP conditions for liquid H$_2$O. Below the HHB, H$_2$O is found in increasingly dense phases spanning liquid, vapour, super-critical, and ice states depending on the location of the HHB and the extent of the H$_2$O layer, as shown in Figure~\ref{fig:phase_diagram}. In the case of a mixed H$_2$O-H/He envelope, the HHB is undefined as it corresponds to an extreme case with no pure H$_2$O layer.

\section{Discussion}
\label{sec:discussion}

Our constraints on the interior and atmospheric properties of K2-18b provide insights into its physical conditions, origins, and potential habitability.  

\subsection{Possible Compositions and Origins} 
\label{sec:scenarios}
Here we discuss three representative classes that span the range of possible compositions, as indicated in Figures \ref{fig:mr_diagram}, \ref{fig:ternary} and \ref{fig:phase_diagram}. The specific cases chosen here fit the $M_{\rm p}$ and $R_{\rm p}$ exactly, as shown in figure \ref{fig:mr_diagram}. A wider range of solutions exist in each of these classes within the 1$\sigma$ uncertainties.

{\it Case 1: Rocky World.} One possible scenario is a massive rocky interior overlaid by a H/He envelope. For example, a pure Fe core of 94.7\%  by mass with an almost maximal H/He envelope of 5\% explains the data with minimal $x_{\rm H_2O}=0.3\%$, consistent with our retrieved H$_2$O abundance in the atmosphere. The HHB in this case is at $\sim 10^{6}$ bar, yielding supercritical H$_2$O close to the ice X phase. It is also possible in this case that the H$_2$O and H/He are mixed, meaning the HHB is not well-defined. Such a scenario is consistent with either H$_2$ outgassing from the interior \citep{elkins-tanton2008,rogers2010} or accretion of an H$_2$-rich envelope during formation \citep{lee2016}.

{\it Case 2: Mini-Neptune.} There are a range of plausible compositions consisting of a non-negligible H/He envelope in addition to significant H$_2$O and core mass fractions, akin to canonical models for Neptune and Uranus \citep{guillot2014}. One such example is a 45\% Earth-like core with $x_{\rm env }=0.03\%$ and $x_{\rm H_2O}=54.97\%$. In this case the HHB is at $P_{\rm HHB} = 700$bar and $T_{\rm HHB}=1500$K, with H$_2$O in the supercritical phase. 

{\it Case 3: Water World.} A $\sim$100\% water world with a minimal H$_2$-rich atmosphere ($x_{\rm env} \sim 10^{-6}$) is permissible by the data. However, such an extreme case is implausible from a planet formation perspective; some amount of rocky core is required to initiate further ice and gas accretion \citep{leger2004, rogers2011,lee2016}. For example, a planet with $x_{\rm core} = 10\%, x_{\rm H_2O} = 89.994\%$ and a thin H/He envelope ($x_{\rm env} = 0.006\%$) can explain the data. For this case, $P_{\rm HHB} = 130$bar and $T_{\rm HHB}=560$K, corresponding to liquid H$_2$O. For the same core fraction, solutions with even smaller H/He envelopes are admissible within the 1$\sigma$ uncertainties on $M_p$ and $R_p$, leading to $P_{\rm HHB}$ and $T_{\rm HHB}$ approaching habitable STP conditions.

\subsection{Potential Habitability} 

A notional definition of habitability argues for a planetary surface with temperatures and pressures conducive to liquid H$_2$O \citep[e.g.,][]{kasting1993,meadows2018}. Living organisms are known to thrive in Earth's extreme environments (extremophiles). Their living conditions span the phase space of liquid H$_2$O up to $\sim$1000 bar pressures at the bottom of the Marianas Trench and $\sim$400 K temperatures near hydrothermal vents \citep[e.g.,][]{merino2019}. 

Whether or not habitable conditions prevail on K2-18b depends on the extent of the H/He envelope. The thermodynamic conditions at the surface of the H$_2$O layer span a wide range in the H$_2$O phase diagram. While most of these solutions lie in the super-critical phase, many others lie in the liquid and vapour phases. Model solutions with core mass fractions $<$15\% and H/He envelopes $\lesssim10^{-3}$ allow for liquid H$_2$O at Earth-like habitable conditions discussed above. One plausible scenario is an ocean world, as discussed in section~\ref{sec:scenarios}, with liquid water approaching STP conditions (e.g., 300 K, $\sim$1-10 bar) underneath a thin H/He atmosphere ($x_{\rm env}\lesssim$10$^{-5}$). 

A number of studies in the past have argued for potential habitability on planets with H/He-rich atmospheres orbiting M Dwarfs \citep[e.g.,][]{pierrehumbert2011,seager2013,koll2019}. Given our constraints  above, we find that K2-18b has a realistic chance of being habitable. Furthermore, our constraints on CH$_4$ and NH$_3$ suggest chemical disequilibrium. Among other possibilities for chemical disequilibrium, e.g. photochemistry, the potential influence of biochemical processes may not be entirely ruled out. Future observations, e.g. with the \textit{James Webb Space Telescope}, will have the potential to refine our findings. We argue that planets such as K2-18b can indeed have the potential to approach habitable conditions and searches for biosignatures should not necessarily be restricted to smaller rocky planets.

\acknowledgments
N.M., M.N., A.P., R.B. acknowledge support from the UK Science and Technology Facilities Council (STFC). L.W. thanks the Gates Cambridge Trust for support toward his doctoral studies.  We thank the anonymous reviewer for their helpful comments on the manuscript. This research is made open access thanks to the Bill \& Melinda Gates foundation.

\bibliographystyle{aasjournal}
% \bibliography{references}

\end{document}